# Survey and Comparison of Optical Switch Fabrication Techniques and Architectures

Ravinder Yadav, Rinkle Rani Aggarwal

Department of Computer Science and Engineering, Thapar University, Patiala

Punjab, India

**Abstract**—The main issue in the optical transmission is switching speed. The optical packet switching faces many significant challenges in processing and buffering. The generalized multilevel protocol switching seeks to eliminate the asynchronous transfer mode and synchronous optical network layer, hence the implementation of IP over WDM (wave length division multiplexing). Optical burst switching attempts to minimize the need for processing and buffering by aggregating flow of data packets in to burst [1]. In this paper there is an extensive overview on current technologies and techniques concerning optical switching.

**Index Terms**—Augmented data vertex, Extinction ratio, Electro optic switches, MEMS

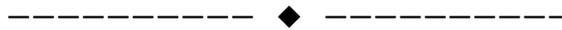

## 1 Introduction

Today's tremendous data transmission demand can't be fulfilled by the electronic transmissions. To fulfill such huge demand optical transmission technology is required. Such system can send 10 to 100 wavelengths per fiber with each wavelength modulated at 10Gbps or more [3]. Before optical all the demand was fulfilled by electronics and now it is considered to be a mature technology that has been studied extensively [1]. But with the growing tremendous demand, optical technology is solution, which depends upon optical switching to fully exploit capacity of optical medium.

The main focus of this paper is on the optical switching which make enable routing of optical signals without conversion of O/E and hence make it data rate and protocols independent.

The paper outlines as follows. In section 2, major functions and applications of optical switches have been presented, in section 3, major issues and parameters in switch fabrications have been discussed. Section 4 presents switch fabrication techniques and conclusions have been presented in section 5.

## 2 Major Functions and Applications of Optical Switches

Optical switches are mainly used in establishing the light path. Following are the major responsibilities of the optical switches expected. Optical cross connect (OXC): OXC are meant to provide a light path that connects two networks nodes [2], [4]. The switches in OXC make them enable to configure a new path [6]. Optical Switch requirements for OXC includes [1]

    a.   Scalability.

    b.   Highly reliable switching.

    c.   Switching without disturbing others.

### 2.1 Protection

In an optical network, sometimes, a single point of failure causes the whole network break down. The protection includes involvement of steps in order to find nature and origin of failure to notify other nodes. So the protection switching is the transmission of data in the event of system or network fault.

### 2.2 Optical Add/Drop multiplexing

The Optical switches must have the capability of addition and deletion of the wave channels without any electronics processing for the high performance. Switches, which bear these capabilities, are called wavelength selective switches.

### 2.3 Optical spectral monitoring

Optical spectral monitoring is an network management operations. In the optical spectral monitoring we receive a small portion of optically tapped signal, separates it into wavelength and monitor for power level, wavelength accuracy and optical cross talk etc.

## 3 Major Issues and Parameters in Switch Fabrications

The major issue in the switch fabrication is the time. Different applications have the different time constraints requirements. Other parameters that should be taken in account are following [1].

### 3.1 Insertion loss

It is measure in DB. It is the fraction of signal power loss due to switch. It is recommended as low as possible.

### 3.2 Cross talk

It is the ratio of power at desired output from desired input to the all other Inputs[1].

### 3.3 Extinction ratio

It is the ratio of output power In ON state to OFF state. It is recommend as large as possible.

Other parameters are reliability, Energy uses and temperature resistance etc.

## 4 Optical Switch Fabrication Techniques

Optical switch fabrication techniques that are in current use are as follows.

### 4.1 Opto-mechanical switches

This was the first commercially used technology for the optical switch fabrication. In this technology switching function is performed by some mechanical means i.e. by the use of prism, mirror, directional coupler etc. To control the switch electrical control signals are used [11]. The major disadvantage is the lack of scalability. Long-term reliability is also of some concern because of mechanical components in Opto-mechanical switches.

### 4.2 Micro electromechanical system devices

These devices are mainly used in the telecom industries. It is a kind of opto-mechanical device.

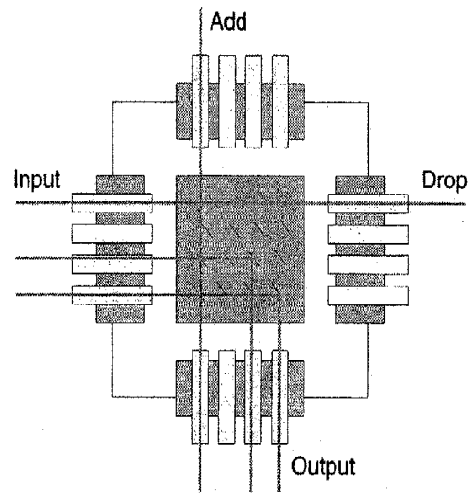

Fig. 1. MEMS switch [1].

Micro-Electro-Mechanical Systems (MEMS) is the integration of mechanical elements, sensors, actuators, and electronics on a common silicon substrate through micro fabrication technology. The micromechanical components are fabricated using compatible "micromachining" processes that selectively etch away parts of the silicon wafer or add new structural layers to form the mechanical and electromechanical devices. MEMS bring together silicon-based microelectronics with micromachining technology, making possible the realization of complete systems-on-a-chip[12]. MEMS switches are mainly of two kinds 2D and 3D. In 2-D MEMS, the switches are digital, since the mirror position is bistable (ON or OFF), which makes driving the switch very straightforward. Figure 1 shows a top view of a 2-D MEMS device with the microscopic mirrors arranged in a crossbar configuration to obtain cross-connect functionality. Collimated light beams propagate parallel to the substrate plane. When a mirror is activated, it moves into the path of the beam and directs the light to one of the outputs, since it makes a 45 angle with the beam. This arrangement also allows light to be passed through the matrix without hitting a mirror. This additional functionality can be used for adding or dropping optical channels (wavelengths). In 3-D MEMS, there is a dedicated movable mirror for each input and each output port. A connection path is established by bending two mirrors independently to direct the light from an input port to a selected output port. Mirrors operate in an analog mode, bending freely about two axes [1]. This is a most promising technology for very-large-port-count OXC switches.

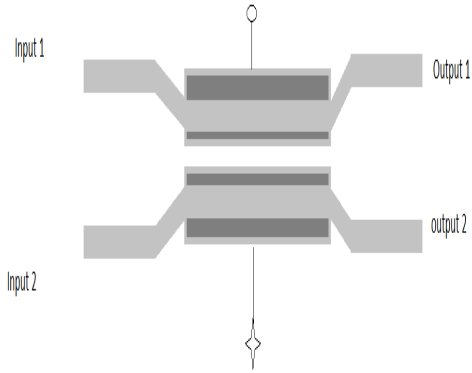

Fig. 2. An electronic directional coupler switch [1]

### 4.3 Electro-optic switches

An 2x2 Electro-optic switch use both electric and optical processing. Electro-optic switch uses a directional coupler whose coupling ratio is changed by varying the refractive index of the material in the coupling region [5]. Commonly used material in the coupling region is lithium niobate (LiNbO) [1]. A switch constructed on a lithium niobate waveguide is shown in figure 2 An electrical voltage applied to the electrodes changes the substrate's index of refraction. The change in the index of refraction manipulates the light through the appropriate waveguide path to the desired port. An Electro-optic switch is capable of changing its state extremely rapidly, typically in less than a nanosecond. This switching time limit is determined by the capacitance of the electrode configuration. Larger switches can be realized by integrating several 2x2 switches on a single substrate. However, they tend to have a relatively high loss and PDL and are more expensive than mechanical switches.

### 4.4 Semiconductor optical amplifier switch

An SOA can be used as an ON–OFF switch by varying the bias voltage. If the bias voltage is reduced, no population inversion is achieved, and the device absorbs input signals. If the bias voltage is present, it amplifies the input signals. It is similar to electrically controlled gate. Figure 3 shows top view of controlled gate. Figure 4 shows a top view of semiconductor. SOA are poor scalable, bear high cost of scalability. It is also difficult to make it PDL independent while constructing a large port count, its fabrication cost grow rapidly.

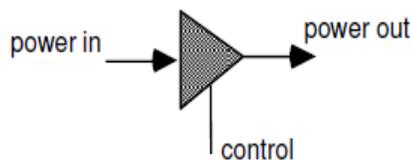

Fig. 3. SOA electrical control gate [13]

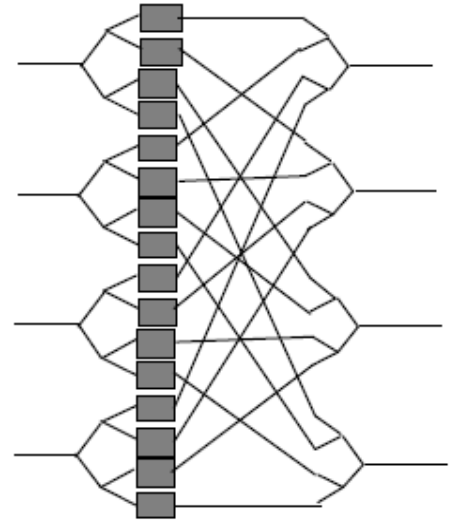

Fig. 4. SOA switch [13]

SOA switching speed is of the order of neon second and that enable it to operates up to 10 Gbps or more data rate at a wave length of 1530nm[2],[14].

All Data in tables have been taken from commercially available datasheets [11]-[18].

## 5 Conclusion

The current electronic network capacity can't accommodate the future market demand. For the high-speed transmission demand of future we have to migrate towards optical technology. In the current trend the capacity of optical medium is not fully exploited. For the different network applications different switches are available. When time is the major constraint, SOA is the best choice, while for high port count MEMS is the best choice. But at the current technology the optical switches are partially electronic, so they can't exploit the full capacity of medium (optical fiber). Purely optical switches needed to be design to fully exploit the capacity of medium and to cope up the future data transmission demand, major technical difficulties need to be overcome [8]-[9].

**Table 1**

Comparison of Optical Switching Technology

| Parameter | Opto-Mechanical Switches | MEMS Switches (8x8) | Electro-optic Switches | Semiconductor Optic Amplifier Switches(PSW-4401) |
|---|---|---|---|---|
| Wavelength (nm) | 1310,1550 both(1310,1550) | 1280~1340 1520~1625 | 1280~1340 1520~1625 | 1530 |
| Insertion loss (DB) | Max 0.60 | Max 3.5 | Max 2.0 | Typ. 0 |
| Return loss (DB) | Min 55 | Min 50 | Max 0.45 | |
| PDL (DB) | Max 0.05 | Max 0.1 | Max 0.1 | Max 2.5 |
| Cross talk (DB) | Min 60 | Max -50 | Max 0.80 | Max 0.80 |
| Switching time | 4ms | 12ms | 5 ns | 3ns |
| Power handling | 1000(mW) | Max 20 (dBm) | 320+10mA 320-10 mA | Max 120mA |
| Operating Temperature | 0~70 C | -5~70 C | 0.5~70 C | 0~40 C |
| Available Configurations | 1x2 2x2 | 8x8 16x16 32x32 | 1x8 1x16 | 1x4 2x1 4x4 |
| Applications | Very low port count applications, Wave length selective application | Large Port Count | Switching time constraint application | Multicasting and broad casting, Strictly non blocking operations |
| Effect on scalability | Increase in Fabrication cost, Reduction in performance | Relatively low increase in cost, Practically size greater than 32x32 is not possible | High losses and PDL. | Increase in fabrication cost, Difficult to make PDL independent. |

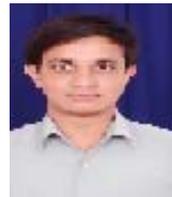

**Ravinder Yadav** received the B.E. degrees in Computer Science and Engineering from SITM in 2005. During 2005 to 2007 have been working in industries. During 2007 to 2008 working as a lecture in DAV Engineering College (Mohindergarh). After 2008 to now persuing ME. In Software from Thapar University, Patiala.

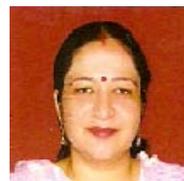

**Rinkle Rani Aggarwal,** B.Tech (Computer Science & Engg.), M.S. (Software Systems), is Assistant Professor in Computer Science & Engineering Department at Thapar University, Patiala.

She has more than 12 years of teaching experience and served academic institutions such as Guru Nanak Dev Engineering College, Ludhiana and S.S.I.E.T 'Derabassi. She has supervised 15 M.Tech. Dissertations and contributed 23 articles in Conferences and 15 papers in research Journals. Her areas of interest are Parallel Computing and Algorithms.